\documentclass[a4paper,11pt,reqno]{article}
\usepackage[dvipsnames]{xcolor}
\usepackage{amsmath}
\usepackage{amsthm}
\usepackage{amssymb}
\usepackage{mathcomp}
\usepackage{dsfont}
\usepackage{comment}
\usepackage{authblk}

\usepackage[active]{srcltx}
\usepackage{stackengine}
\usepackage[bookmarksnumbered=true]{hyperref}

\usepackage{geometry}
\newtheorem{definition}{Definition}[section]

\newtheorem{proposition}[definition]{Proposition}
\newtheorem{theorem}[definition]{Theorem}
\newtheorem{remark}[definition]{Remark}

\numberwithin{equation}{section}

\usepackage[deletedmarkup=xout,authormarkup=none,
]{changes}

\definechangesauthor[name={Ema}, color={brown}]{Ema}

\definechangesauthor[name={Marcello}, color={Emerald}]{Marcello}

\definechangesauthor[name={Marco}, color={BrickRed}]{Marco}

\definechangesauthor[name={Chris}, color={green}]{Chris}

\begin{document}

\title{Bogoliubov theory for the dilute Fermi gas in three dimensions}
\author{Emanuela L. Giacomelli}
\affil{LMU M\"unich, Department of Mathematics, Theresienstr. 39, 80333 M\"unchen, Germany}
%
%
\maketitle

\abstract{In a dilute system of $N$ fermions with spin $1/2$ in three dimensions, we study the correlation energy which is given by the  difference between the ground state energy and the energy of the non-interacting Fermi sea state. We review some recent results from the paper by Falconi \textit{et al.} [Ann. Henri Poincaré 22 (2021), 2283-2353], where we make use of the almost-bosonic nature of the low-energy excitations of the system which can be effectively described by Bogoliubov theory.}

\section{Introduction}\label{sec:int}

The analysis of many body quantum systems of interacting particles has represented a very rich line of research for several decades. In particular, the understanding of interacting Fermi gases is an important problem in mathematical physics.

In this paper we consider the low density Fermi gas which is a dilute system introduced by Gell-Mann and Brueckner \cite{GMB} in 1957 to model nuclear matter; nowadays it plays an important role in the study of cold atomic gases. 

More precisely, we take into account $N$ interacting spin 1/2 fermions in a box $\Lambda_L := [0,L]^3$ with periodic boundary conditions. We study the ground state energy of the system, in the thermodynamic limit, in the case in which the interaction potential is positive and of short-range. We want to characterize the correlation energy of the system, given by the difference between the many-body ground state energy and the energy of the free (non interacting) Fermi sea. We will achieve this goal by making a rigorous connection from Fermi gases to Bose gases and employs Bogoliubov approximation for the later.

In his seminal paper in 1947 Bogoliubov introduced an approximation method which has become the basis of the mathematical theory of interacting Bose gases \cite{Bog}. Since then, many works were devoted to the rigorous justification of the Bogoliubov theory from first principles. For the ground state energy, we refer to the work of \cite{LS1comp, LS2comp, Sol} on one and two components Bose gases and to \cite{YY,FS,BaCS} on the Lee Huang Yang formula in the thermodynamic limit (see also \cite{FGJMO} for the corresponding two dimensional case). In the fixed volume setting the Bogoliubov excitation spectrum has been justified in the mean field regime \cite{S,GS, LNSS, DN, Piz, BSP} as well as in the GP regime \cite{BBCS2, NT, BSS, HST}. See also \cite{ABS, BCS, DG} for results on Bose gases in intermediate regimes between GP and the thermodynamic limit.

In contrast to the flourishing recent developments in the theory of bosonic systems, there are not many mathematical works devoted to the understanding of the collective behavior of fermionic systems. In this setting, the interesting problem is to go beyond the validity of the Hartree-Fock approximation for the ground state energy, which was established in a number of seminal works by Bach \cite{BA}, Graf and Solovej \cite{GSol} (see also references therein), and to derive an asymptotic for the correlation energy. Very recently, rigorous results appeared in this direction. The exact leading order of the correlation energy was derived for the high density Fermi gas by Benedikter et al. \cite{BNPSS} and by Christiansen et al. \cite{CHN} (see also \cite{BNPSS0, BPSS, BNPSS1}). For the low density regime, the best known result goes to Lieb, Seiringer and Solovej in 2005 (see \cite{LSS}), where they proved the asymptotic formula\footnote{Here we use units such that $\hbar = 1$ and we set the masses of the particles to be equal to $1/2$.}
\begin{equation}\label{eq: energy density asymptotics}
    e_L(\rho_\uparrow, \rho_\downarrow) = \frac{3}{5}(6\pi^2)^{\frac{2}{3}}\big(\rho^{\frac{5}{3}}_\uparrow + \rho^{\frac{5}{3}}_\downarrow \big) + 8\pi a \rho_\uparrow\rho_\downarrow + o(\rho^2),
\end{equation}
for the ground state energy (with $\rho_\sigma$ being the density of the particles with spin $\sigma= \uparrow, \downarrow$). 

Recently in \cite{FGHP} we offered a new approach to study the ground state energy of the low density Fermi gas, in particular we gave an alternative proof of \eqref{eq: energy density asymptotics} using a bosonization technique and Bogoliubov approximation. We will review this approach below.

We set the total number of particles to be $N = N_\uparrow + N_\downarrow$, where in general $N_\sigma$ denotes the number of particles with a given spin $\sigma = \uparrow, \downarrow$. The Hamiltonian describing the system is given by 
\[
    H_N = -\sum_{i=1}^N \Delta_{x_i} + \sum_{i<j = 1}^N V(x_i - x_j),
\]
where we denote by $\Delta_{x_i}$ the Laplacian acting on the $i$-th particle, and $V$ is a pair interaction potential defined as 
\[
    V(x-y) = \frac{1}{L^3}\sum_{p\in \frac{2\pi}{L}\mathbb{Z}^3} e^{ip\cdot (x-y)}\hat{V}_\infty(p),
\]
with $\hat{V}_\infty = \int_{\mathbb{R}^3}\, dx e^{-ip\cdot x} V_\infty(x)$. We suppose $V_\infty$ to be compactly supported and smooth. Since we work with fermionic particles and being the Hamiltonian be spin independent, $H_N$ acts on the antisymmetric Hilbert space given by $L^{2}_a(\Lambda_L^{N_\uparrow})\otimes N L^2_a (\Lambda_L^{N_\downarrow})$, where $L^2_a(\Lambda_L^{N_\sigma}) = L^2(\Lambda_L)^{\wedge N_\sigma}$ is the antisymmetric sector of $L^2(\Lambda_L)^{\otimes N_\sigma}$. We can then define the ground state energy of the system as 
\[
    E_L(N_\uparrow, N_\downarrow) = \inf_{\Psi\in L^{2}_a(\Lambda_L^{N_\uparrow})\otimes N L^2_a (\Lambda_L^{N_\downarrow})}\frac{\langle\Psi, H_N \Psi\rangle}{\langle \Psi, \Psi\rangle}.
\]
Let $\rho = \rho_\uparrow + \rho_\downarrow$ be the total density of the system (recall that $\rho_\sigma$ denotes the density of particles with spin $\sigma = \uparrow, \downarrow$, i.e., $\rho_\sigma = N_\sigma/L^3$). Being $H_N$ translation invariant, we can define the ground state energy density as 
\[
    e_L(\rho_\uparrow, \rho_\downarrow) := \frac{E_L(N_\uparrow, N_\downarrow)}{L^3}.
\]
We are interested in studying $e_L(\rho_\uparrow, \rho_\downarrow)$ in the thermodynamic limit, i.e., in the limit $N_\sigma, L \rightarrow \infty$ with $\rho_\sigma$ fixed. It is well know that this limit exists and that it is independent of the boundary conditions on $\Lambda_L$ \cite{Ro, Ru}. Moreover, we focus in the dilute regime, which corresponds to the condition $\rho \ll 1$. The main result we want to review here reads as follows.

\begin{theorem}\label{teo: main} Let $V\in C^\infty (\Lambda_L)$, compactly supported, $V\geq 0$. There exists $L_0>0$ large enough such that for $L \geq L_0$ the following holds:
\begin{equation}
    e_L(\rho_\uparrow, \rho_\downarrow) = \frac{3}{5}\big(6\pi^2\big)^{\frac{2}{3}} \big(\rho_\uparrow^{\frac{5}{3}} + \rho_\downarrow^{\frac{5}{3}}\big) + 8\pi a \rho_\uparrow\rho_\downarrow + r_L(\rho_\uparrow, \rho_\downarrow),
\end{equation}
where $a$ is the scattering length of the potential $V$, and for some constant $C$ only dependent on $V$:
\begin{equation}
    -C \rho^{2+\xi_2} \leq r_L (\rho_\uparrow, \rho_\downarrow) \leq C\rho^{2+\xi_1},
\end{equation}
with $\xi_1 = 2/9$ and $\xi_2 = 1/9$.
\end{theorem}
With respect to the proof proposed in \cite{LSS}, we are able to improve the error estimates; however, we need to restrict our analysis to very regular potentials and, in particular we cannot take into account the hard-core case as in \cite{LSS}. 

Our proof gives a different point of view on the dilute Fermi gas, via a rigorous implementation of the ideas of Bogoliubov theory for Fermi systems. Making use of Bogoliubov theory, we can indeed effectively describe the low energy excitations around the free Fermi gas as pairs of fermions, which can be though as emergent bosonic particles. 

We think that this method can be developed further to derive more refined energy asymptotics. In this framework, an outstanding open problem is to rigorously prove the Huang and Yang (HY) formula proposed in \cite{HY}, in which the next-order correction to the ground state energy of $3D$ dilute fermions is of the order $\rho^{7/3}$. More precisely, we expect that the techniques developed in the proof of Theorem \ref{teo: main}, paves the way to the proof of the (HY) formula which in units such that $\hbar = 1$ and setting the masses of the particles to be equal to $1/2$ reads as:
\begin{multline}
    e_L(\rho_\uparrow, \rho_\downarrow) = \frac{3}{5}(6\pi^2)^{\frac{2}{3}}\big(\rho^{\frac{5}{3}}_\uparrow + \rho^{\frac{5}{3}}_\downarrow \big) + 8\pi a \rho_\uparrow\rho_\downarrow
    \\
     +\frac{4(11 -2\log 2)}{35\pi^2} \left(\frac{3}{4\pi}\right)^{4/3} a^2 \rho^{7/3} + o(\rho^{7/3})
\end{multline}

\noindent\textbf{Outline of the paper.} Th{}e paper is organizes as follows. In Section \ref{sec: free fermi gas} we calculate the energy of the free Fermi gas which gives the first order correction to the energy density. In section \ref{sec: particle-hole} we isolate the correlation energy of the Fermi system using the well know particle-hole transformation. In Section \ref{sec: bogoliubov} we give the main ideas for the proof of Theorem \ref{teo: main}. In particular, we discuss how to implement Bogoliubov theory in our fermionic setting. \\

\noindent\textbf{Acknowledgments.} The author acknowledges the support of the Istituto Nazionale di Alta Matematica ``F. Severi", through the Intensive Period ``INdAM Quantum Meetings (IQM22)".

\section{The free Fermi gas}\label{sec: free fermi gas}
We refer to the free Fermi gas (FFG) as to the Slater determinant that minimizes the kinetic energy of the system. More precisely, this state can be written as 
\begin{multline}
    \Psi_{\mathrm{FFG}}\big(\{x_i, \uparrow\}_{i=1}^N,, \{y_j, \downarrow\}_{j=1}^{N_\downarrow}\big)
    \\
    = \frac{1}{\sqrt{N_\uparrow !}}\frac{1}{\sqrt{N_\downarrow !}} \big(\mathrm{det} f^\uparrow_{k_i}(x_j)\big)_{1\leq i,j\leq N_\uparrow}\big(\mathrm{det} f^\downarrow_{k_i}(x_j)\big)_{1\leq i,j\leq N_\downarrow},
\end{multline}
with
\[
    f^\sigma_k(x) \equiv f_k(x) = \frac{e^{ik\cdot x}}{L^{\frac{3}{2}}}, \qquad \mbox{for}\,\,\, k\in \mathcal{B}^\sigma_F,
\]
where $\mathcal{B}^{\sigma}_F$ denotes the Fermi ball:
\[
    \mathcal{B}^\sigma_F := \bigg\{ k \in \frac{2\pi}{L} \mathbb{Z}^3\, \big \vert\, |k| \leq k_F^\sigma\bigg\}, \qquad k_F^\sigma \,(\equiv \mbox{Fermi}\,\, \mbox{momentum}) = (6\pi^2)^{\frac{1}{3}} \rho_\sigma^{\frac{1}{3}} + o_L(1).
\]
Note that there cannot be repetition in the momenta appearing in the definition of $\Psi_{\mathrm{FFG}}$, this is due to the Pauli's principle. Thus, $\Psi_{\mathrm{FFG}}$ is the most uncorrelated state one could work with in a fermionic system.
\begin{remark}[Completely filled Fermi ball] We only take into account values $N_\sigma$ such that $N_\sigma = | \mathcal{B}_{F}^\sigma |$, i.e., we suppose that the Fermi ball is completely filled. This is enough for our purposes since the densities $\rho_\sigma = N_\sigma/L^3$ we get with these values $N_\sigma$ are dense in $\mathbb{R}_+$ as $L\rightarrow + \infty$.
\end{remark}
By some direct calculations, it is possible to show that the energy of the free Fermi gas is given by the Hartree-Fock energy function, i.e.,  
\begin{multline}\label{eq: HF energy}
    \langle \Psi_{\mathrm{FFG}}, H_N \Psi_{\mathrm{FFG}}\rangle = E_{\mathrm{HF}}
    \\
    = \sum_{\sigma}\sum_{k\in\mathcal{B}^\sigma_F} |k|^2  + \frac{L^3}{2}\sum_{\sigma, \sigma^\prime}\hat{V}(0) \rho_\sigma \rho_{\sigma^\prime} - \frac{1}{2}\sum_{\sigma, \sigma^\prime}\frac{\delta_{\sigma, \sigma^\prime}}{L^3}\sum_{k,k^\prime\in \mathcal{B}^\sigma_F} \hat{V}(k-k^\prime).
\end{multline}
\begin{remark}[$\Psi_{\mathrm{FFG}}$ - quasi free state] It is possible to write $E_{\mathrm{HF}}$ with respect to the reduced one-particle density matrix of $\Psi_{\mathrm{FFG}}$, i.e., 
\begin{equation}\label{eq: def w}
    \omega_{\sigma, \sigma^\prime}(x;x^\prime) := \delta_{\sigma,\sigma^\prime} \frac{1}{L^3}\sum_{k\in\mathcal{B}^\sigma_F} e^{ik\cdot (x-x^\prime)}.
\end{equation}
More precisely, 
\begin{multline}\label{eq: HF energy}
E_{\mathrm{HF}} \equiv E_{\mathrm{HF}}(\omega) 
\\
= - \mathrm{tr}\Delta\omega + \frac{1}{2} \sum_{\sigma, \sigma^\prime} \int_{\Lambda_L \times \Lambda_L} \hspace{-0.3cm}dxdy\, V(x-y) \big(\omega_{\sigma, \sigma}(x;x)\omega_{\sigma^\prime, \sigma^\prime}(y;y) - |\omega_{\sigma, \sigma^\prime}(x;y)|^2\big),
\end{multline}
Thus $\Psi_{\mathrm{FFG}}$ is an example of quasi free state.

\end{remark}

Doing a Taylor expansion and using the symmetry of the interaction potential $V$, we can write 
\begin{equation}\label{eq: taylor approx HF}
    -\delta_{\sigma, \sigma^\prime}\frac{1}{2L^3}\sum_{k,k^\prime\in\mathcal{B}^\sigma_F}\hat{V}(k-k^\prime) = -\frac{L^3}{2}\sum_\sigma\hat{V}(0)\rho_\sigma^2 + \mathcal{O}(L^3\rho^{\frac{8}{3}}).
\end{equation}
Inserting then \eqref{eq: taylor approx HF} in \eqref{eq: HF energy}, we get that 
\begin{equation}\label{eq: HF per volume}
    \frac{\langle \Psi_{\mathrm{FFG}}, H_N \Psi_{\mathrm{FFG}}\rangle}{L^3} = \frac{3}{5}(6\pi^2)^{\frac{2}{3}}\big(\rho_\uparrow^{\frac{5}{3}} + \rho_\downarrow^{\frac{5}{3}}\big) + \hat{V}(0)\rho_\uparrow\rho_\downarrow + \mathcal{O}(\rho^{\frac{8}{3}}) + \mathcal{O}(L^{-1}),
\end{equation}
where the last error is due to the approximation of sums with integrals. For more details we refer to \cite[Section 3.2.1]{FGHP}.
\begin{remark}[The role of correlations] 
Comparing \eqref{eq: HF per volume} with $\eqref{eq: energy density asymptotics}$, it is clear that the free Fermi gas gives the right order asymptotics (up to the second term) with the wrong constant, i.e., we have $\hat{V}(0) = \int dx V(x)$ in place of $8\pi a$. This means that restricting our analysis to $\Psi_{\mathrm{FFG}}$ we are missing the correlations between the particles in the many-body ground state.
\end{remark} 
\section{The particle-hole transformation}\label{sec: particle-hole}{}
From the previous section it is clear that to prove an asymptotic of the form \eqref{eq: energy density asymptotics}, we have to take into account the correlations between particles. The first step in this direction is to be able to compare the many body ground state energy with the energy of the free Fermi gas. This can be done by introducing a unitary transformation, usually know as \textit{particle-hole transformation}. To do that, it is convenient to work in second quantization. We then introduce the fermionic Fock space as
\begin{equation}
    \mathcal{F} = \bigoplus_{n\geq 0} \mathcal{F}^{(n)}, \qquad \mathcal{F}^{(n)} = L^2(\Lambda_L; \mathbb{C}^2)^{\wedge n},
\end{equation}
where we set $\mathcal{F}^{(0)} = \mathbb{C}$. To denote a given element $\psi\in \mathcal{F}$, we write $\psi = (\psi^{(0)},\psi^{(1)}, \cdots , \psi^{(n)}, \cdots )$. We will refer to $\Omega = (1, 0, \cdots, 0, \cdots)$ as to the vacuum, i.e., the zero particle state. The Hamiltonian in second quantization is 
\begin{multline}
    \mathcal{H} = \sum_{\sigma}\int_{\Lambda_L} dx\, \nabla_x a^\ast_{x,\sigma} \nabla_x a_{x,\sigma}
    \\
     + \frac{1}{2}\sum_{\sigma, \sigma^\prime}\int_{\Lambda_L \times \Lambda_L}dxdy\, V(x-y) a^\ast_{x,\sigma}a^\ast_{y,\sigma^\prime} a_{y,\sigma^\prime}a_{x,\sigma},
\end{multline}
where $a^\ast_{x,\sigma}, a_{x,\sigma}$ are the usual creation (resp. annihilation) operator-valued distributions. We omit here all the details (see \cite[Section 3.1]{FGHP}). Before proceeding further, we just recall the definition of the creation (resp. annihilation) operators in momentum space. We have 
\begin{equation}
    \hat{a}_{k,\sigma} = \frac{1}{L^{\frac{3}{2}}}\int_{\Lambda_L} dx\, a_{x,\sigma}e^{-ik\cdot x}, \qquad \hat{a}^\ast_{k,\sigma} = (\hat{a}_{k,\sigma})^\ast.
\end{equation}
Moreover, the number operator is defined as 
\begin{equation}
    \mathcal{N} = \sum_\sigma\sum_{k\in\frac{2\pi}{L}\mathbb{Z}^3} \hat{a}^\ast_{k,\sigma}\hat{a}_{k,\sigma} = \sum_\sigma \int_{\Lambda_L}dx\, a^\ast_{x,\sigma}a_{x,\sigma}.
\end{equation}
Note that $\mathcal{N}$ counts the number of particles in a given sector of the fermionic Fock space, i.e., $(\mathcal{N}\psi)^{(n)} = n \psi^{(n)}$.

Since we are interested in studying $N= N_\uparrow + N_\downarrow$ fermionic particles, if we set $\mathcal{F}^{(N_\uparrow, N_\downarrow)}$ to be the subspace of the Fock space such that for any $\psi \in \mathcal{F}^{(N_\uparrow, N_\downarrow)}$ one has $\mathcal{N}\psi = N \psi = (N_\uparrow + N_\downarrow)\psi$, we can rewrite the ground state energy as 
\begin{equation}
    E_L(N_\uparrow, N_\downarrow) = \inf_{\psi \in\mathcal{F}^{(N_\uparrow, N_\downarrow)}} \frac{\langle \psi, \mathcal{H}\psi\rangle}{\langle \psi, \psi\rangle}.
\end{equation}
We can now introduce the particle-hole transformation. 
\begin{definition}[Particle-hole transformation $R$]\label{def: particle-hole transf} Let $\omega_{\sigma, \sigma^\prime}$ be the reduced one-particle density matrix of the free Fermi gas defined in \eqref{eq: def w}. Let $u,v:L^{2}(\Lambda_L; \mathbb{C}^2)\rightarrow L^2(\Lambda_L; \mathbb{C}^2)$ be the operators having integral kernel given by
\begin{equation}
    u_{\sigma,\sigma^\prime}(x;y) := \delta_{\sigma,\sigma^\prime} \delta(x-y) - \omega_{\sigma,\sigma^\prime}(x;y), \qquad v_{\sigma,\sigma^\prime}(x;y) = \delta_{\sigma,\sigma^\prime}\sum_{k\in\mathcal{B}^\sigma_F}|\overline{f_k}\rangle \langle f_k|,
\end{equation}
where $\delta(x-y)$ denotes the periodic Delta distribution on $\Lambda_L$. The \textit{particle-hole transformation} $R$ is defined has the unitary operator $R: \mathcal{F}\rightarrow \mathcal{F}$ such that the following holds:
\begin{itemize}
    \item[$\mathrm{(i)}$] The vector $R\Omega$ is an $N-$ particle state such that 
    \begin{equation}
        (R\Omega)^{(n)} = 0 \quad \mbox{for}\,\,\,n\neq N, \qquad (R\Omega)^{(N)} = \Psi_{\mathrm{FFG}}.
    \end{equation}
    \item[$\mathrm{(ii)}$]\, The map $R:\mathcal{F}\rightarrow \mathcal{F}$ is such that 
    \begin{equation}\label{eq: conjugation under R}
        R^\ast a_{x,\sigma} R = a_\sigma(u_x) + a^\ast_\sigma (\overline{v}_x),
    \end{equation}
    where we set $u_{\sigma,\sigma}\equiv u_\sigma$, $v_{\sigma,\sigma} \equiv v_\sigma$, $u_x(y)\equiv u(x;y)$, $v_x(y)\equiv v(x;y)$ and 
    \begin{equation}
        a_\sigma(u_x) = \int\, dy\, \overline{u}_\sigma(y;x) a_{y,\sigma}\, \qquad a^\ast_\sigma(\overline{v}_x) = \int\, dy\, \overline{v}_\sigma(y;x) a^\ast_{y,\sigma}
    \end{equation}
\end{itemize}
\end{definition}
The existence of the unitary $R$ defined above is ensured by the Shale-Stinespring theorem, and in general $R$ is a Bogoliubov transformation (we refer to \cite{Sol} for more details).
\begin{remark}[Orthogonality relation $u,v$] It is clear from the definition of $u$ and $v$ that 
\[
    u\overline{v} = 0, \qquad \overline{v}v = \omega.
\]
More precisely, the operator $u$ is a projection outside the Fermi sea and $\omega$ projects inside the Fermi ball. Looking at the relation \eqref{eq: conjugation under R} one can then expect that conjugating the Hamiltonian via $R$ it is possible to extract the contribution coming from the Fermi sea from the energy.
\end{remark}
\begin{remark}[Action of $R$ in momentum space] Using that $\hat{a}_{k,\sigma} = a_\sigma(f_k)$, an equivalent formulation of \eqref{eq: conjugation under R} is the following one:
\begin{equation}
    R^\ast \hat{a}_{k,\sigma}R = \begin{cases}  \hat{a}_{k,\sigma} &\mbox{if}\,\,\, k\not\in\mathcal{B}_F^\sigma, \\ \hat{a}^\ast_{k,\sigma} &\mbox{if}\,\,\, k\in \mathcal{B}_F^\sigma,   \end{cases}
\end{equation}
i.e., conjugating via $R$ we are creating holes inside the Fermi sea and particles outside $\mathcal{B}_F^\sigma$.
\end{remark}
Conjugating the Hamiltonian $\mathcal{H}$ via the particle-hole transformation $R$, one gets the following result.
\begin{proposition}[Conjugation via $R$]\label{pro: conjug R} Let $\psi \in \mathcal{F}$ be a normalized state, such that $\langle \psi, \mathcal{N}_\sigma, \psi \rangle = N_\sigma$ and such that $N = N_\uparrow + N_\downarrow$. Then, it holds true: 
\begin{equation}
    \langle \psi, \mathcal{H}\psi\rangle = E_{\mathrm{HF}}(\omega) + \langle R^\ast\psi, \mathbb{H}_0 R^\ast \psi\rangle + \langle  R^\ast \psi, \mathbb{X} R^\ast \psi\rangle + \sum_{i=1}^4\langle R^\ast \psi, \mathbb{Q}_i R^\ast \psi \rangle, 
\end{equation}
where the operators $\mathbb{H}_0$, $\mathbb{X}$, $\mathbb{Q}_i$ $\mathrm{(}i=1, \ldots,4\mathrm{)}$ are defined as follows:
\begin{equation}
    \mathbb{H}_0  = \sum_{k,\sigma}||k|^2 - (k_F^\sigma)^2| \hat{a}_{k,\sigma}^\ast \hat{a}_{k,\sigma},
\end{equation}
\begin{equation}
    \mathbb{X} = \int\, dxdy\, V(x-y) \omega_\sigma(x;y) \big(a^\ast_\sigma(u_x)a_\sigma(u_y) - a^\ast_\sigma(\overline{v}_y)a_\sigma(\overline{v}_x)\big),
\end{equation}
\begin{equation}
    \mathbb{Q}_1 = \frac{1}{2}\sum_{\sigma, \sigma^\prime} \int\, dxdy\, V(x-y) a^\ast_\sigma(u_x)a^\ast_{\sigma^\prime}(u_y)a_{\sigma^\prime}(u_y)a_\sigma(u_x),
\end{equation}  
\begin{eqnarray}
    \mathbb{Q}_2 &=& \frac{1}{2}\sum_{\sigma, \sigma^\prime}\int\, dxdy\, V(x-y)\, \big( a_\sigma^\ast(u_x)a^\ast_\sigma(\overline{v}_x)a_{\sigma^\prime}(\overline{v}_y) a_{\sigma^\prime}(u_y)\\
    && \qquad  - 2a^\ast_\sigma(u_x) a^\ast_{\sigma^\prime}(\overline{v}_y) a_{\sigma^\prime}(\overline{v}_x) a_\sigma(u_x) \nonumber + a^\ast_{\sigma^\prime}(\overline{v}_y)a^\ast_\sigma(\overline{v}_x)a_\sigma(\overline{v}_x)a_{\sigma^\prime}(\overline{v}_y) \big),
\end{eqnarray}  
\begin{eqnarray}
    \mathbb{Q}_3 &=& - \sum_{\sigma, \sigma^\prime} \int\, dxdy\, V(x-y) \big(a^\ast_\sigma(u_x)a^\ast_{\sigma^\prime}(u_y)a^\ast_{\sigma}(\overline{v}_x)a_{\sigma^\prime}(u_y) 
    \\
    &&\qquad \qquad - a^\ast_{\sigma}(u_x)a^\ast_{\sigma^\prime}(\overline{v}_y) a^\ast_\sigma(\overline{v}_x) a_{\sigma^\prime}(\overline{v}_y) \big) + \mathrm{h.c.},\nonumber
\end{eqnarray}
\begin{equation}
    \mathbb{Q}_4 = \frac{1}{2}\sum_{\sigma, \sigma^\prime} \int\, dxdy\, V(x-y) a^\ast_{\sigma}(u_x) a^\ast_{\sigma^\prime}(u_y) a^\ast_{\sigma^\prime}(\overline{v}_y) a_\sigma^\ast(\overline{v}_x) + \mathrm{h.c.}
\end{equation}
\end{proposition}
For the proof of Proposition \ref{pro: conjug R} we refer to \cite[Proposition 3.1]{FGHP}.
\begin{remark}[Neglecting the interaction of equal spin particles]
In what follows, using the positivity of the interaction potential $V$, we can discard (in a lower bound) the interaction of particles with equal spin, i.e., under the same hypothesis of Proposition \ref{pro: conjug R} one can also prove that 
\begin{equation}
    \langle\psi, \mathcal{H}\psi\rangle \geq E_{\mathrm{HF}}(\omega) + \langle R^\ast\psi, \mathbb{H}_0 R^\ast \psi\rangle + \langle R^\ast \psi, \mathbb{X} R^\ast \psi \rangle + \langle R^\ast \psi, \widetilde{\mathbb{Q}} R^\ast \psi \rangle,
\end{equation}
where each $\widetilde{\mathbb{Q}}_i$, $i= 1, \ldots, 4$ is defined as the corresponding $\mathbb{Q}_i$ neglecting the equal spin case, i.e., replacing the sum over $\sigma, \sigma^\prime$ with the sum over $\sigma \neq \sigma^\prime$.
\end{remark}
We now state some a priori estimates which are useful to neglect all the quantities which are not contributing to the first order asymptotics for the correlation energy. To do that, we restrict our analysis to those states which are close enough to the ground state energy. To quantify the meaning of \textit{close enough} we give then the following definition.
\begin{definition}[Approximate ground state]\label{def: approx ground state}
Let $\psi \in \mathcal{F}$ be a normalized state, such that $\langle \psi, \mathcal{N}_\sigma\psi\rangle = N_\sigma$, with $N = N_\uparrow + N_\downarrow$. We say that $\psi$ is an approximation ground state if 
\begin{equation}
    \left|\langle \psi, \mathcal{H}\psi\rangle - \sum_{\sigma}\sum_{k\in\mathcal{B}_F^\sigma} |k|^2\right| \leq CL^3 \rho^2.
\end{equation}  
\end{definition}
 Following \cite[Section 3]{FGHP}, it is possible to rigorously prove what follows. 
\begin{proposition}\label{pro: a priori bounds} Let $\psi$ be an approximate ground state. Then 
\begin{equation}
    \langle R^\ast\psi,\mathbb{H}_0 R^\ast\psi\rangle \leq CL^3\rho^2, \quad \langle R^\ast \psi, \mathbb{Q}_1 R^\ast \psi \rangle \leq CL^3\rho^2, \quad \langle R^\ast \psi, \mathbb{Q}_1 R^\ast \psi \rangle \leq CL^3\rho^2.
    \end{equation}
Moreover, it also follows that
    \begin{equation}
        \langle R^\ast \psi, \mathcal{N} R^\ast \psi \rangle \leq CL^3\rho^{\frac{7}{6}}.
\end{equation}
\end{proposition} 
For the proof of Proposition \ref{pro: a priori bounds}, we refer to \cite[Lemma 3.5, Lemma 3.6, Corollary 3.7, Lemma 3.9]{FGHP}.
As a consequence of the bounds in Proposition \ref{pro: a priori bounds}, one can easily prove (see \cite[Proposition 3.3]{FGHP}) that if $\psi$ is an approximate ground state, then 
\begin{equation}
   |\langle R^\ast\psi, \mathbb{X}R^\ast\psi\rangle|,\,|\langle R^\ast\psi, \mathbb{Q}_2R^\ast\psi\rangle|,\, |\langle R^\ast\psi, \widetilde{\mathbb{Q}}_2R^\ast\psi\rangle| \leq C\langle R^\ast\psi, \mathcal{N} R^\ast \psi\rangle \leq CL^3\rho^{2+\frac{1}{6}}.
\end{equation}
   Moreover, it also holds that 
\begin{equation}
    |\langle \psi, \mathbb{Q}_3 \psi \rangle |, \, |\langle \psi, \widetilde{\mathbb{Q}}_3 \psi \rangle | \leq C L^3 \rho^{2+\frac{1}{12}}.
\end{equation}
Note that the bound above can be improved up to $\rho^{2+\frac{1}{9}}$ using a better estimate for the number operator $\mathcal{N}$ (see Proposition \ref{pro: improved N}).
We can conclude this section stating that if $\psi$ is an approximate ground state then
\begin{equation}\label{eq: correl en H0 Q1 Q4}
 \langle R^\ast \psi, (\mathbb{H}_0 + \widetilde{\mathbb{Q}}_1 + \widetilde{\mathbb{Q}}_4) R^\ast \psi\rangle + \mathcal{E}_{\mathrm{lb}} \leq \langle \psi, \mathcal{H}\psi\rangle - E_{\mathrm{HF}} \leq \langle R^\ast \psi, (\mathbb{H}_0 + \mathbb{Q}_1 + \mathbb{Q}_4) R^\ast \psi\rangle + \mathcal{E}_{\mathrm{ub}},
\end{equation}
where both $\mathrm{\mathcal{E}}_{\mathrm{ub}}$ and  $\mathrm{\mathcal{E}}_{\mathrm{lb}}$ are sub-leading with respect to $L^3\rho^{2}$. 

\section{Bogoliubov Theory for Fermi systems}\label{sec: bogoliubov}

 In what follows we recall the main steps useful to derive the first order correction for the correlation energy, i.e. to prove Theorem \ref{teo: main}. The starting point is the estimate in \eqref{eq: correl en H0 Q1 Q4}: if $\psi$ is an approximate ground state, then 
\begin{equation}\label{eq: bog 1}
    \langle \psi, \mathcal{H}\psi\rangle \simeq E_{\mathrm{HF}}+ \langle R^\ast \psi, (\mathbb{H}_0 + \widetilde{\mathbb{Q}}_1 + \widetilde{\mathbb{Q}}_4) R^\ast \psi\rangle.
\end{equation}
Note that with respect to \eqref{eq: correl en H0 Q1 Q4}, we are ignoring the contribution coming from equal spin. In the lower bound this is automatically true using the positivity of the interaction as explained above; for the upper bound, this can be done thanks to the choice of the trial state (see \cite[Section 7]{FGHP}).

From now on we ignore all the errors in the approximations we discuss; for the rigorous proof we refer to \cite{FGHP}. As already underlined, the idea is to make use of the almost-bosonic nature of the low energy excitations  around the Fermi sea to describe the correlation energy: this can be done taking into account pairs of fermions which behaves as effective bosons. To this intent, we introduce the \textit{pseudo-bosonic} operators:
\begin{equation}
    \hat{b}_{p,\sigma} = \int\, dx\, e^{ip\cdot x} a_\sigma(u_x)a_\sigma(\overline{v}_x) = \sum_{k}\hat{u}(k+p)\hat{v}(k) \hat{a}_{k+p,\sigma}\hat{a}_{k,\sigma},\,\, \hat{b}^\ast_{p,\sigma} = \big(\hat{b}_{p,\sigma}\big)^\ast 
\end{equation}
where $\hat{u}$ and $\hat{v}$ are the indicator functions of $(\mathcal{B}_F^\sigma)^c$ and $\mathcal{B}_F^\sigma$, respectively, i.e., 
\begin{equation}
    \hat{u}(k) = \begin{cases} 1 &\mbox{for}\,\, k\notin \mathcal{B}_F^\sigma, \\ 0 &\mbox{otherwise},\end{cases} \qquad \hat{v}(k) = \begin{cases} 1 &\mbox{if}\,\, k \in \mathcal{B}_F^\sigma, \\ 0 &\mbox{otherwise}.  \end{cases}
\end{equation}
It holds that the operators $\mathbb{H}_0$, $\mathbb{Q}_1 $ and $\mathbb{Q}_4$ can be written, up to some errors, with respect to the $\hat{b}, \hat{b}^\ast$ operators. More precisely, when acting over an approximate ground state $\psi$, one can approximate (see \cite[Section 4]{FGHP}):
\[
    \mathbb{H}_0 \psi\simeq \frac{1}{L^3}\sum_\sigma \sum_p |p|^2 \hat{b}_{p,\sigma}^\ast\hat{b}_{p,\sigma}\psi, 
\]
\[
    \widetilde{\mathbb{Q}}_1 \psi\simeq \frac{1}{2L^3}\sum_{\sigma\neq\sigma^\prime}\sum_{p,q,r} \hat{V}(p)\hat{b}^\ast_{q-p,\sigma}\hat{b}_{q,\sigma}\hat{b}^\ast_{r-p,\sigma^\prime}\hat{b}_{r,\sigma} - \frac{1}{2L^3}\sum_\sigma\sum_{p,q}\hat{V}(p)\hat{b}_{q,\sigma}^\ast \hat{b}_{q,\sigma}\psi,
\]
The term $\widetilde{\mathbb{Q}}_4$ instead is a bosonizable one, i.e., 
\[
    \mathbb{Q}_4 = \frac{1}{2L^3}\sum_{\sigma\neq \sigma} \sum_p \hat{V}(p)\hat{b}_{p,\sigma}\hat{b}_{-p,\sigma} + \mathrm{h.c.}
\]
The idea then is to use a pseudo Bogoliubov transformation to diagonalize $\mathbb{H}_0 + \widetilde{\mathbb{Q}}_1 + \widetilde{\mathbb{Q}}_4$ to leading order. 
\\

\noindent\textbf{Pseudo bosonic Bogoliubov transformation.} We start by introducing all the quantities which will appear in the definition of the Bogoliubov transformation. In particular, we need to regularize the operators $u,v$ introduced in Definition \ref{def: particle-hole transf}. This is due to the fact that we work in the thermodynamic limit, and then we need to do some localization in order to get some decay of the kernel of the operators $u$ and $v$. This is the role of the cut-off defined below. We set 
\begin{equation}
    v^r_{\sigma, \sigma^\prime} = \frac{\delta_{\sigma, \sigma^\prime}}{L^3}\sum_k \hat{v}^r_\sigma(k) |\overline{f}_k\rangle \langle f_k|, \qquad u^r_{\sigma, \sigma^\prime} = \frac{\delta_{\sigma, \sigma^\prime}}{L^3} \sum_k \hat{u}^r_\sigma(k)|f_k\rangle \langle f_k|,
\end{equation} 
where $\hat{v}^r_\sigma$, $\hat{u}^r_\sigma$ are smooth, radial functions such that 
\begin{equation}\label{eq: def cut off}
    \hat{v}^r_\sigma(k) = \begin{cases} 1 &\mbox{for}\,\, |k| < k_F^\sigma - \rho_\sigma^\alpha,   \\ 0 &\mbox{for}\,\, |k| \geq k_F^\sigma,\end{cases} \qquad \hat{u}^r_\sigma(k) = \begin{cases} 0 &\mbox{for}\, \, |k| \leq k_F^\sigma, \\ 1 &\mbox{for}\,\, 2k_f^\sigma \leq |k| \leq \frac{3}{2}\rho^{-\beta}_\sigma, \\ 0 &\mbox{for}\,\, |k| \geq 2\rho^{-\beta}_\sigma, \end{cases}
\end{equation}
for some $\alpha > 1/3$, $\beta >0$.

Recall now that we want to define a Bogoliubov transformation useful to extract the quantity $8\pi a \rho_\uparrow\rho_\downarrow$ from the correlation energy, where $a$ is the scattering length associated to the interaction potential. It is then natural to expect that the solution of the scattering equation will play an important role in the definition of the pseudo-bosonic Bogoliubov transformation. We define $\varphi_\gamma$ to be the solution of the scattering equation in a ball $B\equiv B_{\rho^{-\gamma}}(0)\subset \mathbb{R}^3$ centered at zero and with radius $\rho^{-\gamma}$, satisfying Neumann boundary conditions, i.e., $\varphi_\gamma$ is such that 
\begin{equation}
    -\Delta(1-\varphi_\gamma) + \frac{1}{2}V_\infty (1-\varphi_\gamma) = \lambda_\gamma (1-\varphi_\gamma), \quad \varphi_\gamma = \nabla \varphi_\gamma = 0 \, \, \mbox{on}\,\, \partial B,
\end{equation}
where $|\lambda_\gamma| \leq C\rho^{3\gamma}.$
We briefly recall here that 
\begin{equation}\label{eq: 8pi a}
    8\pi a_\gamma = \int\, dx\, V(x)(1-\varphi_\gamma (x)) = 8\pi a + \mathcal{O}(\rho^\gamma),
\end{equation}
for this and other properties of $\varphi_\gamma$ we refer to \cite[Appendix A]{FGHP}. We denote by $ \widetilde{\varphi}_\gamma$ is the periodization of $\varphi_\gamma$ over $\Lambda_L$. 

\begin{definition}[Pseudo bosonic Bogoliubov transformation] Let $\gamma \geq 0$, $\lambda \in [0,1]$. We define the unitary operator $T_\lambda: \mathcal{F}\rightarrow \mathcal{F}$ as: 
\begin{equation}
    T_\lambda := e^{\lambda(B-B^\ast)}, \quad B:= \int\, dzdz^\prime \widetilde{\varphi}_\gamma(z-z^\prime) b_{z,\uparrow}^r b_{z^\prime, \downarrow}^r,
\end{equation}
where we set $b_{\cdot,\sigma}^r = a_\sigma(u_{\cdot}^r)a_\sigma(\overline{v}^r_{\cdot})$. 
Moroever, we set $T\equiv T_1$.\\
\end{definition}
\begin{remark}
The reason to use the scattering equation in a ball with Neumann boundary condition is due to the fact that we need to localize $\widetilde{\varphi}_\gamma$ in order to have some decay estimates useful to bound several error terms.
\end{remark}

\noindent\textbf{Useful bounds.} An important tool in order to rigorously prove Theorem \ref{teo: main} is the propagation of the a priori bounds for $\mathcal{N}$, $\mathbb{H}_0$, $\mathbb{Q}_1$ stated in Proposition \ref{pro: a priori bounds}. 

\begin{proposition}[Propagation of the estimates for $\mathbb{H}_0$, $\mathbb{Q}_1$]\label{pro: propagation est}
    Let $\lambda\in [0,1]$, let $5/18 \leq \gamma \leq 1/3$. If $\psi$ is an approximate ground state, it holds
    \[
        \langle T^\ast_\lambda R^\ast \psi, \mathbb{H}_0 T^\ast_\lambda R^\ast \psi\rangle \leq CL^3\rho^2, \quad \langle T^\ast_\lambda R^\ast \psi, \mathbb{Q}_1 T^\ast_\lambda R^\ast \psi \rangle \leq CL^3\rho^2.
    \]
    Moreover, using that $\widetilde{\mathbb{Q}}_1 \leq \mathbb{Q}_1$, it holds 
    \[
        \langle T^\ast_\lambda R^\ast \psi, \widetilde{\mathbb{Q}}_1 T^\ast_\lambda R^\ast \psi \rangle \leq CL^3\rho^{2}.
    \]
\end{proposition}
For the proof of Proposition \ref{pro: propagation est} we refer to \cite[Section 5]{FGHP}. We also mention an improved estimate for the number operator $\mathcal{N}$ with respect to the one in stated in Section \ref{sec: particle-hole}. 
\begin{proposition}[Improved a priori estimate for $\mathcal{N}$]\label{pro: improved N} Let $\lambda\in [0,1]$ and let $\gamma \leq 1/2$. If $\psi$ is an approximate ground state, it holds
\begin{equation}
    \langle \xi_\lambda, \mathcal{N}\xi_\lambda \rangle \leq CL^{\frac{3}{2}}\rho^{\frac{1}{6}}\|\mathbb{H}_0^{\frac{1}{2}} T^\ast R^\ast \psi\| + CL^3\rho^{2-\gamma}.
\end{equation}
\end{proposition}
The proof can be found in \cite[Section 5]{FGHP}. 
\\

\noindent\textbf{Ideas for the proof of Theorem \ref{teo: main}.} 
We have now all the ingredients to summarize the main steps for the proof of Theorem \ref{teo: main}. Here, we only discuss the ideas for the proof of the lower bound, being the upper bound a simple adaptation. We focus on the operators $\mathbb{H}_0 + \widetilde{\mathbb{Q}}_1 + \widetilde{\mathbb{Q}}_4$. We then use Duhamel formula to write
\begin{multline*}
    \langle R^\ast \psi, (\mathbb{H}_0 + \widetilde{\mathbb{Q}}_1 + \widetilde{\mathbb{Q}}_4) R^\ast \psi \rangle
    \\
    = \langle T^\ast R^\ast \psi, (\mathbb{H}_0 + \widetilde{\mathbb{Q}}_1) T^\ast R^\ast \psi\rangle - \int_0^1\, d\lambda\, \frac{d}{d\lambda}\langle T^\ast_\lambda R^\ast \psi, (\mathbb{H}_0 + \widetilde{\mathbb{Q}}_1)T^\ast_\lambda R^\ast \psi\rangle
    \\
    + \langle T^\ast R^\ast \psi, \widetilde{\mathbb{Q}}_4 T^\ast R^\ast \psi\rangle - \int_0^1\, d\lambda\, \frac{d}{d\lambda}\langle T^\ast_\lambda R^\ast \psi, \widetilde{\mathbb{Q}}_4 T^\ast_\lambda R^\ast \psi\rangle
\end{multline*} 
We first look at the integral in the second line above, it holds true that (see \cite{FGHP}):
\begin{multline}\label{eq: comm H0 + Q1}
    \int_0^1 d\lambda\,\frac{d}{d\lambda} \langle T^\ast_\lambda R^\ast \psi, (\mathbb{H}_0 + \mathbb{Q}_1) T^\ast_\lambda R^\ast \psi \rangle = \int_0^1\, d\lambda\, \langle T^\ast_\lambda R^\ast \psi, [\mathbb{H}_0 + \mathbb{Q}_1, B-B^\ast] T^\ast_\lambda R^\ast \psi\rangle
    \\
    \sim \int_0^1\, d\lambda\int dxdy\, \big(2\Delta\widetilde{\varphi}_\gamma (x-y) - V(x-y)\widetilde{\varphi}_\gamma(x-y)\big) \langle T^\ast_\lambda R^\ast \psi, b^r_{x,\uparrow} b^r_{y, \downarrow} T^\ast_\lambda R^\ast \psi \rangle + \mathrm{c.c.}
\end{multline}
Here, we underline that to estimate the error terms we ignored above, one needs to use the properties of the scattering solution as well as the bounds in Proposition \ref{pro: propagation est} and Proposition \ref{pro: improved N}. Moreover, one has also to deal with the regularization induced by the cut-off in the Bogoliubov transformation, this is not a trivial task (see e.g., \cite[Proposition 5.5, Lemma 5.6]{FGHP}). Using now that
\[
    2\Delta\varphi(x) + V(x)(1-\varphi(x)) = \lambda_\gamma (1-\varphi(x))\qquad \mbox{on} \,\,B_{\rho^{-\gamma}}(0),
\]
we get, denoting by $\chi_{_{B_{\rho^{-\gamma}}}}$ the indicator function of the ball $B_{\rho^{-\gamma}}$, that 
\begin{multline}\label{eq: comm H0 + Q1}
    \int_0^1 d\lambda\,\frac{d}{d\lambda} \langle T^\ast_\lambda R^\ast \psi, (\mathbb{H}_0 + \mathbb{Q}_1) T^\ast_\lambda R^\ast \psi \rangle = \int_0^1\, d\lambda\, \langle T^\ast_\lambda R^\ast \psi, [\mathbb{H}_0 + \mathbb{Q}_1, B-B^\ast] T^\ast_\lambda R^\ast \psi\rangle
    \\
    \sim \int_0^1 d\lambda \int dxdy\, \chi_{_{B_{\rho^{-\gamma}}}}(x-y)\big( - V(x-y) + \lambda_\gamma (1 -\widetilde{\varphi}_\gamma(x-y))\big) \cdot \\
    \cdot \langle T^\ast_\lambda R^\ast \psi, b^r_{x,\uparrow} b^r_{y,\downarrow} T^\ast_\lambda R^\ast \psi \rangle + \mathrm{c.c.}
    \\
    \sim -\int_0^1 d\lambda\int dxdy\, \langle T^\ast_\lambda R^\ast \psi, \mathbb{Q}^r_4 T^\ast_\lambda R^\ast \psi \rangle, 
\end{multline}
where we introduce a regularized version of the bosonizable operator $\mathbb{Q}_4$, i.e., 
\begin{equation}
    \widetilde{\mathbb{Q}}_4^r = \int\, dxdy\, V(x-y) b_{x,\uparrow}^r b_{\downarrow,y}^r+ \mathrm{h.c.},
    .
\end{equation}
We also stress that the error coming from the scattering equation, i.e., from the operator $ \lambda_\gamma (1 -\widetilde{\varphi}_\gamma(x-y))\big) b^r_{x,\uparrow}b^r_{y,\downarrow}$ is $o(L^3\rho^2)$ (here one has to use again Proposition \ref{pro: propagation est} and Proposition \ref{pro: improved N}, see \cite[Proposition 5.7]{FGHP})
One can then prove that for any approximate ground state, 
\begin{equation}\label{eq: reg Q4}
    \langle T^\ast R^\ast \psi, \mathbb{Q}_4 T^\ast R^\ast\psi \rangle \sim \langle T^\ast R^\ast\psi , \mathbb{Q}_4^r T^\ast R^\ast\psi \rangle.
\end{equation}
Note that to rigorously justify the approximation above, one has to carefully choose the cut-off in $\hat{u}^r$ and $\hat{v}^r$ in order to have a sub-leading contribution with respect to $\rho^2$. Combining \eqref{eq: comm H0 + Q1} and \eqref{eq: reg Q4} and using Duhamel's formula again, we then get 
\begin{multline}
 {}   - \int_0^1\, d\lambda\, \frac{d}{d\lambda}\langle T^\ast_\lambda R^\ast \psi, (\mathbb{H}_0 + \widetilde{\mathbb{Q}}_1)T^\ast_\lambda R^\ast \psi\rangle
    + \langle T^\ast R^\ast \psi, \widetilde{\mathbb{Q}}_4 T^\ast R^\ast \psi\rangle
    \\
    \sim  \int_0^1 d\lambda\int_\lambda^1 d\lambda^\prime\, \frac{d}{d\lambda^\prime}\langle T^\ast_\lambda R^\ast \psi, \widetilde{\mathbb{Q}}_4^r T^\ast_\lambda R^\ast \psi\rangle - \int_0^1 d\lambda \frac{d}{d\lambda}\langle T^\ast R^\ast \psi, \widetilde{\mathbb{Q}}_4^r T^\ast R^\ast \psi \rangle.
\end{multline}
From the last line above, we can extract the constant term proportional to $\rho^2L^3$ and contributing to first order correction of the correlation energy asymptotics. More precisely, following \cite[Proposition 6.1]{FGHP}, one can prove that 
\begin{equation}\label{eq: }
    \frac{d}{d\lambda} \langle T^\ast_\lambda R^\ast \psi, \widetilde{\mathbb{Q}}_4^r T^\ast_\lambda R^\ast \psi\rangle \sim 2\rho^{r}_\uparrow \rho^r_\downarrow\int\, dxdy\, V(x-y)\widetilde{\varphi}_\gamma(x-y),
\end{equation}
where $\rho_\sigma^r$ denotes the regularized density of the particles with spin $\sigma$. The last thing to do is then to compare the regularized densities with the true ones: this can be done up to errors which depend on the cut-off over $\hat{v}$. One can prove indeed that $|\rho_\sigma - \rho_\sigma^r| \leq C\rho^{2/3 + \alpha}$, where $\alpha >0$ is as in \eqref{eq: def cut off}. In particular, it follows that, for any $\alpha >1/3$, the product $\rho_\uparrow^r \rho_\downarrow^r$ can be replaced by $\rho_\uparrow\rho_\downarrow$ up to errors $o(L^3\rho^2)$.
Combining then all the estimates together, one gets 
\begin{multline}
    \langle R^\ast(\mathbb{H}_0 + \widetilde{\mathbb{Q}}_1 + \widetilde{\mathbb{Q}}_4) R^\ast \psi \rangle 
    \\
    \sim C \langle T^\ast R^\ast \psi, (\mathbb{H}_0 + \widetilde{\mathbb{Q}}_1) T^\ast R^\ast \psi \rangle - 2\rho_\uparrow \rho_\downarrow \int\, dxdy V(x-y)\widetilde{\varphi}_\gamma (x-y),  
\end{multline}
for some constant $C<1$ (this is due to the fact that we need to use the positive operators $\mathbb{H}_0$, $\widetilde{\mathbb{Q}}_1$ to bound several error terms).
Inserting the asymptotics above in \eqref{eq: bog 1}, we then have
\begin{multline}\label{eq: final}
    \frac{1}{L^3}\langle \psi \mathcal{H}\psi \rangle \sim \frac{E_{\mathrm{HF}}}{L^3} - \frac{2}{L^3}\rho_\uparrow \rho_\downarrow \int\, dxdy V(x-y)\widetilde{\varphi}_\gamma (x-y)
    \\
    \sim \frac{3}{5}(6\pi^2)^{\frac{2}{3}} (\rho_\uparrow^{\frac{5}{3}} + \rho_\downarrow^{\frac{5}{3}}) + 8\pi a \rho_\uparrow \rho_\downarrow,
\end{multline}
where we used \eqref{eq: HF per volume} together with \eqref{eq: 8pi a}. Note that above we discarded the positive contribution $C\langle T^\ast R^\ast \psi, (\mathbb{H}_0 + \widetilde{\mathbb{Q}}_1) T^\ast R^\ast \psi \rangle$. This indeed can be ignored in the proof of the lower bound, being positive, and it is vanishing for the upper bound, thanks to the choice of the $N-$particle trial state $RT\Omega$. To conclude, we write the optimal choices for the parameters appearing in our analysis. In the lower bound, after some optimization, we fix $\gamma=1/3$ and we take the value $\alpha$ appearing in \eqref{eq: def cut off} to be $2/3$: with this choices the error we commit in \eqref{eq: final} of the order $\rho^{2+\frac{1}{9}}$, as stated in Theorem \ref{teo: main}. For the upper bound instead, we take $\gamma = 2/9$ and $\alpha =2/3 + 1/42$, producing a final error of the order $\rho^{2+ 2/9}$ in \eqref{eq: final}.

\end{document}